# Aging research of the LAB-based liquid scintillator in stainless steel container[*]

CHEN Hai-tao[1,2;1)]  YU Bo-xiang[2,3;2)]  SHAN Qing[1]  DING Ya-yun[2,3]

DU Bing[2,3]  Liu Shu-tong[2]  ZHANG Xuan[2,3]  ZHOU Li[2,3]

JIA Wen-bao[1]  FANG Jian[2,3]  Ye Xing-chen[2]  HU Wei[2,3]

Niu Shun-li[2,3]  Yan Jia-qing[2]  Zhao Hang[2,3]  Hong Dao-jin[2]

[1]Nanjing University of Aeronautics and Astronautics, Jiangsu 210016, China

[2]State Key Laboratory of Particle Detection and Electronics, Beijing 100049, China

[3]Institute of High Energy Physics, CAS, Beijing 100049, China

**Abstract:** Stainless steel is the material used for the storage vessels and piping systems of LAB-based liquid scintillator in JUNO experiment. Aging is recognized as one of the main degradation mechanisms affecting the properties of liquid scintillator. LAB-based liquid scintillator aging experiments were carried out in different material of containers (type 316 and 304 stainless steel and glass) at two different temperature (40 and 25 degrees Celsius). For the continuous liquid scintillator properties tests, the light yield and the absorption spectrum are nearly the same as that of the unaged one. The attenuation length of the aged samples is 6%~12% shorter than that of the unaged one. But the concentration of element Fe in the LAB-based liquid scintillator does not show a clear change. So the self aging has small effect on liquid scintillator, as well as the stainless steel impurity quenching. Type 316 and 304 stainless steel can be used as LAB-based liquid scintillator vessel, transportation pipeline material.

## 1. Introduction

Jiangmen Underground Neutrino Observatory (JUNO), whose target-detector will be filled with 20 kton liquid scintillator (LS) of the $3\%/\sqrt{E(MeV)}$ energy resolution, can determine the mass hierarchy at a confidence level of $\Delta x^2_{MH} \sim (10 \div 12)(3 \div 3.5\sigma)$[1]. To measure the energy spectrum of the reactor neutrinos precisely, high energy response and transparency of LS are needed, i.e., JUNO LS has to achieve attenuation length of 22 $m$, light yield of 1200 $pe/MeV$, and radiation background $< 10^{-15}$ $g/g$. Aging is recognized as one of the main degradation mechanisms affecting the properties of LS.

LS used in this study is Daya Bay LS having the same composition with JUNO LS : linear alkyl benzene (LAB) as the solvent, 3 $g/L$ 2,5-diphenyloxazole (PPO) as the fluor, and 15 $mg/L$ p-bis-(o-methylstyryl)-benzene (bis-MSB) as the wavelength shifter[2]. The Luminescence Mechanism of the JUNO LS should be discussed. Particles will lose part or all of their energy when they enter LS, and produce excited solvent molecules. Then solvent molecules emit certain wavelengths of UV or visible light in the process of de-excitation. And they also undergo energy

[*] Supported by National Natural Science Foundation of China (11205183,11005117,11225525,11390384)

1) E-mail: chenht@ihep.ac.cn

2) E-mail: yubx@ihep.ac.cn(corresponding author)

transition when colliding with other particles. These fluorescent photons go through Rayleigh scattering, refraction, reflection process, constantly have been absorbed and emitted. And Rayleigh scattering refers to a process where the light is scattered elastically, without a change in wavelength[3]. Finally, they reach the photocathode of PMT. The quenching effects of the liquid scintillator are caused by Cherenkov light emission and subsequent absorption and re-emission in the LAB-based LS[4–6]. If aging effects like quenching and oxidation presents in LS, the energy transfered from the excited solvent molecule become less due to self absorption effects. In this study, the aging effects were identified by evaluating the light yield, the attenuation length, the absorption spectrum and the change of Fe content.

In addition to considering LS self aging effects, the impurity quenching effects caused by storage vessels should also be considered. Stainless steel is the material for the storage vessels and piping systems of LS in JUNO experiment. Some stainless steel tanks will be used as LS storage vessel as long as three years. Among different stainless steel grades, the 304 and 316 types are widely used. They are generally highly resistant to liquid organics, but this quality depends on the composition and concentration of the liquid organics and the surrounding temperature[7, 8]. So properties of LS need to be tested after long-time contact with 316/304 stainless steel.

Heating method plays an important role in accelerating reaction rate of an experiment. Besides temperature, the properties of LS may also be influenced by the aging time and contact area with the stainless steel container. As normally LS is stored at room temperature, fast aging experiments with less extra high-temperature reaction were conducted at a close temperature (40 ℃). And aging experiments at room temperature (25 ℃) were conducted as a comparison.

## 2. Experimental procedure

## 2.1 Experimental setup

For A LS aging experimental system with two different heat-treatment conditions (40 ℃ and 25 ℃) was conducted. A schematic drawing of the experimental setup (40 ℃) is shown in Fig. 1. Two incubators ((a) and (b)) were used to control temperature with a variation of ±1 ℃. Four glass conical flasks were put in incubator (a). The volume of each flask was 2*L*, and the flasks were sealed well to avoid the quenching and oxidation effects[4–6]. Type 316 and 304 cylindrical stainless steel tanks were placed inside cuboid incubator (b). Each tank had a volume of 15*L* and a diameter of 21 *cm*, with 10*L* LS inside. 600 *ml*/minute nitrogen was blown into each tank and finally into a bubble beaker to avoid the quenching and oxidation effects too. To increase the contact area between LS with stainless steel to accelerate aging speed, a 316 and a 304 stainless steel sheet of 180×430 *mm$^2$* were placed into corresponding tanks. In order to monitor and record the temperature of each LS container, two thermocouple temperature sensors were placed on a glass conical flask and the 316 stainless steel tank. LS in four flasks stabilized at 38.75~40.50 ℃, and LS in 316 and 304 tank stabilized at 39.75~40.50 ℃. We took out 2*L* LS samples from glass conical flasks and type 316 and 304 stainless steel tanks on the 53th, 94th, 144th, 192th day. Aging experiments of 25 ℃ were conducted as a comparison. A 316 and a 304 tank of the same geometry, with 10*L* of LS but without sheet, were put at room temperature (25 ℃). As well, the tanks were sealed well to avoid the quenching and oxidation effects. LS inside stabilized at 25±2 ℃. And we took out a LS sample of 2*L* on the 307th day.

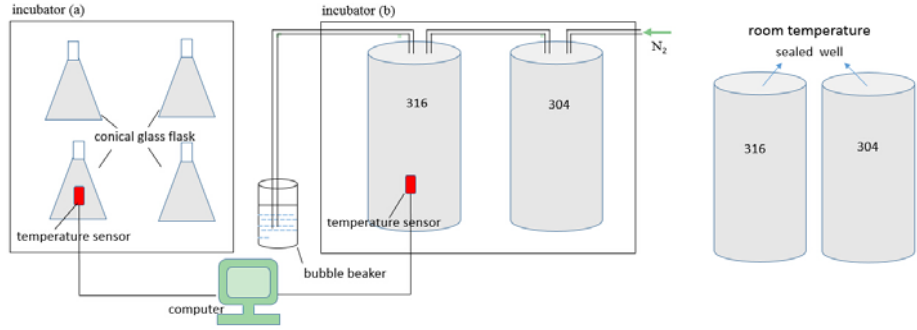

Fig. 1. LS aging system

## 2.2 Aging time conversion for stainless steel tanks

For LS aging experiments were carried out in stainless steel tanks at room temperature and 40 ℃. Different LS samples from steel tanks correspond to different reaction rates, respectively.

According to Van't Hoff[9], for homogeneous thermochemical reaction, the reaction temperature increases 10 ℃, the reaction rate becomes 2~4 times the original reaction rate. Defined as $k = Q^{\frac{T_2-T_1}{10}}$, where $k$ represents the chemical reaction rate, i.e., reaction rate at experimental temperature $T_2$ becomes $Q$ times the $T_1$ temperature.

Comparing 40 ℃ and 25 ℃ aging experiments, adding stainless steel sheets increased the contact area between stainless steel and LS by up to about 1.5 times, so a correction of the reaction rate needs to be done, a factor brought by contact area was added ($\sigma=1.5$), and $Q$ value was taken as 3 (the median value). So, $t_{eq.} = \sigma\, t_{ag.} 3^{\frac{T_2-T_1}{10}}$, which means $t_{ag.}$ time at $T_2$ in the stainless steel tank, the equivalent working time will be $t_{ag.}$ at 25 ℃.

According to the equation, at 40 ℃, 53 days aging time of LS is equivalent to 1.13 years at 25 ℃, 94 days equivalent to 2 years, 144 days equivalent to 3 years, 192 days equivalent to 4.1 years. Time conversion is shown in Fig. 2.

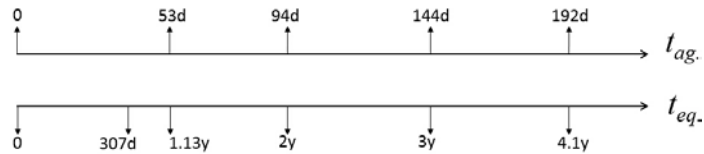

Fig. 2. Time conversion

## 3 Results

### 3.1 Relative light yield measurement

The light yield of the LS is generally measured via Compton scattering of γ rays. When colliding with γ rays, excited LS molecules by ionization will emit certain wavelengths of ultraviolet or visible light. Fig. 3(a) shows the pulse amplitude spectrum without coincidence detection, and it is a Compton platform.

To reduce the fit error, a coincidence detector was added, which consists of a LaBr$_3$ crystal and a photomultiplier tube. Fig. 3(b) shows the pulse amplitude spectrum with coincidence detection, and the peak can be fitted with Gaussian function.

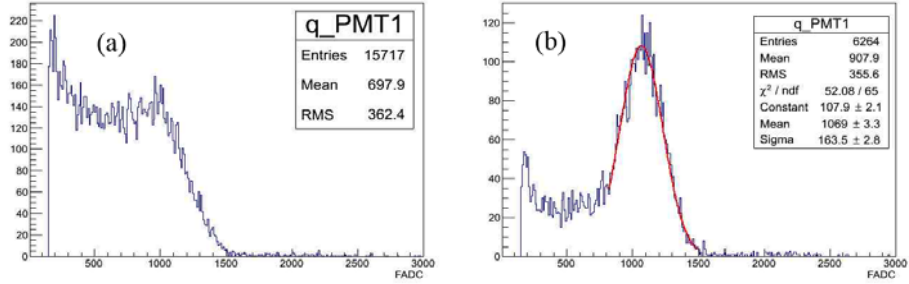

Fig. 3. LS energy spectrum of non-coincidence detection (a) and coincidence detection (b)

As shown in Fig. 4, the LS samples were irradiated by a radioactive source $^{137}$Cs (15 $\mu Ci$). Each LS sample quantitatively weighed 130 $g$. All LS measurements were performed in the dark room.

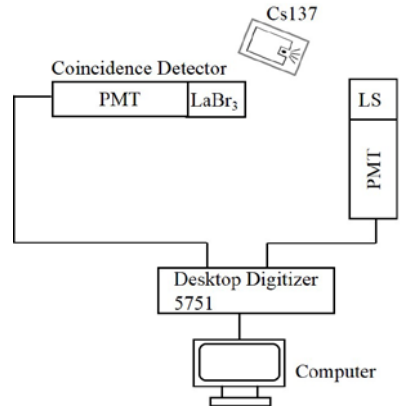

Fig. 4. Experiment setup for measuring the LS light yield

The measurement result shows that the light yield reduces 2% after aging (Fig. 5). In consideration of statistical error, also 2%, the light yield is nearly the same as that of the unaged one. So with long-time contact with the 316 and 304 stainless steel, the light yield of LAB-based LS reduces very little.

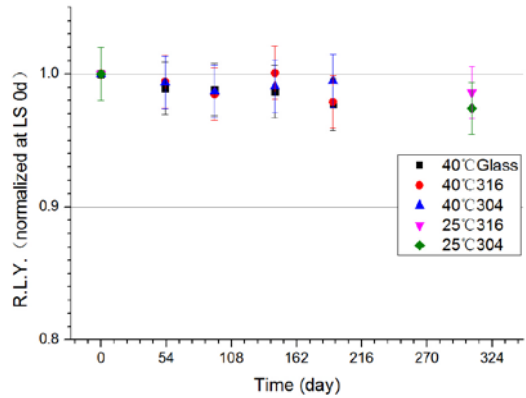

Fig. 5. The result of relative light yield (R.L.Y) measurement

## 3.2 Attenuation length measurement

Attenuation length $L$ is defined as the distance that photons pass through a sample, whose number reduces to 1/e, namely the distance that 63% of the photons are absorbed or scattered[10]. It can be represented by the formula $I(\lambda, x) = I_0(\lambda) e^{-\frac{x}{L}}$, $\lambda$ represents the wavelength of the light beam (430±5 *nm*), $I_0(\lambda)$ represents the initial monochromatic light intensity, *x* represents the height of the LS. So the light intensity $I(\lambda,x)$ that reaches photomultiplier changes with *x*.

Attenuation length measurement device is shown in Fig. 6. A pulse generator was used to drive the lightemitting diode (LED), which was used as the light source. Then light beam went through fiber to the lens, the 430 *nm* filter. Then passed through a $\phi$ 5 *mm* aperture diaphragm to form a bunch of parallel beam. The sample container's length is 1.5 *m*.

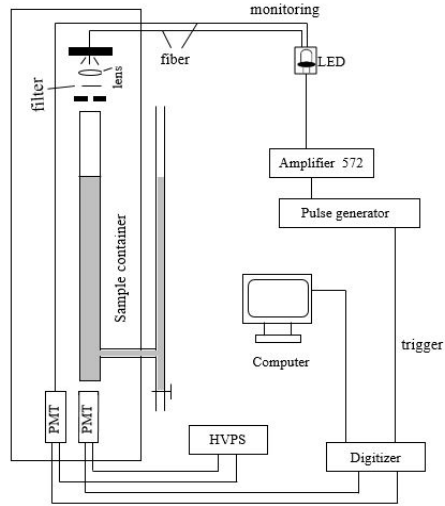

Fig. 6. Attenuation length measurement device

The stability of the measurement system was measured about 200 minutes. When the system was stable, we changed the height of LS to get a set of data. Each time we reduced LS height for 10~15 *cm*, and at least 7 points were measured.

Table 1. Specific data of aging LS's attenuation length

| Aging Time (day) | Attenuation length of LS (m) | | |
|---|---|---|---|
| | Glass | 316 | 304 |
| 0 | 14.84±0.53 | 14.84±0.53 | 14.84±0.53 |
| temperature 40℃ | | | |
| 53 | 14.72±0.69 | 14.18±0.75 | 14.05±0.73 |
| 94 | 14.15±0.85 | 13.92±0.80 | 13.65±0.79 |
| 144 | 14.45±0.68 | 14.28±0.78 | 14.10±0.78 |
| 192 | 13.99±0.87 | 13.39±0.77 | 13.02±0.68 |
| temperature 25℃ | | | |
| 307 | | 14.07±0.68 | 13.89±0.71 |

Specific data is shown in Table 1. Result shows that the attenuation length reduces 6%~12%, and type 316 stainless steel has a relatively smaller impact on LS's attenuation length than 304 (Fig. 7).

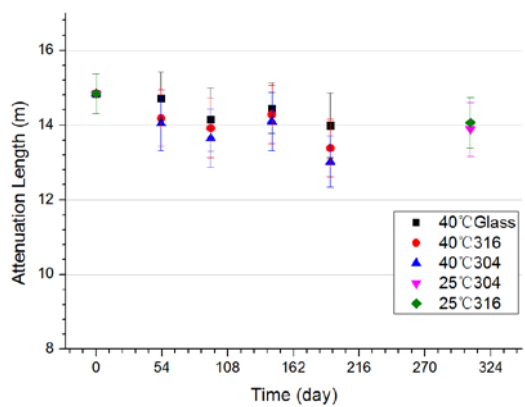

Fig. 7. Attenuation length of aging LS

## 3.3 Absorption spectrum measurement

A L650 UV-Vis spectrophotometer was used to measure LS absorption spectrum corresponding to the wavelength range of 190~900 *nm*. As shown in Fig. 8, absorption spectrum of LS samples in the wavelength range of 410~510 nm is very similar. So, the change of aged LS samples absorption spectrum is very little.

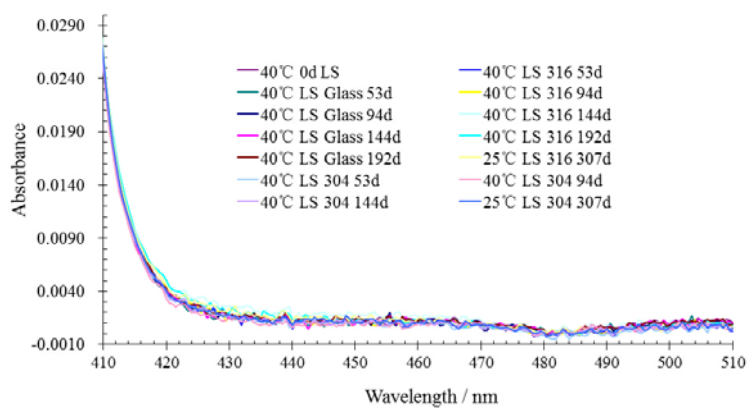

Fig. 8. Absorption spectrum measurement result

## 3.4 Fe concentration measurement

The stainless steel may affect the properties of LS by impurity quenching effects. We analyzed impurities released to LS by examining the Fe impurity. 6 *mol/L* nitric acid was used to extract Fe from organic phase into aqueous phase. Then aqueous phase was diluted and measured by ICP-MS. The result is shown in Table 2. So the concentration of element Fe in LS does not show a clear change when contact with type 316/304 stainless steel.

Table 2. The examined Fe concentration

| Samples of Aged LS | Fe concentration(ppm) |
|---|---|
| 0d | 1.84±0.13 |
| 316type 25℃307d | 1.69±0.11 |
| 316type 40℃192d | 1.90±0.35 |
| 304type 25℃307d | 1.82±0.19 |
| 304type 40℃192d | 1.69±0.11 |

## 4 Conclusions

A set of LS aging experiments have been conducted and a suitable LS aging time calculation method has been applied. 53, 94, 144, 192 days LS aging experiments at 40 ℃ equivalent to 1.13, 2, 3, 4.1 years at room temperature (25 ℃), and 307 days 25 ℃ LS aging experiment has been done as a comparative comparison. The light yield and the absorption spectrum of the aged samples are nearly the same as that of the unaged one. The attenuation length reduces 6%~12%. The concentration of element Fe in LS does not show a clear change.

The results revealed that the self aging has small effect on LS, as well as the stainless steel impurity quenching. 304 or 316 stainless steel can be used as LAB-based LS vessel, transportation pipeline material.

## References

[1] Li Y F, Cao J, Wang Y F, and Zhan L, Phys. Rev. D, 2013,88:013008
[2] Ding Y Y et al., Nucl. Instr. and Meth. A, 2008, 584: 238
[3] G Alimonti et al., Nucl. Instr. and Meth. A, 2000, 440(2): 363
[4] F.P. An et al., Nucl. Instr. and Meth. A, 2012, 685: 88
[5] XIAO H L et al., Chin. Phys. C, 2010, 34(05): 571
[6] LI X B et al., Chin. Phys. C, 2011, 35(11): 1026
[7] A.J. Invernizzi et al., Materials Science and Engineering: A, 2008, 485(1-2):234
[8] X.L Cheng et al., Corrosion Science, 1998, 41(2):321
[9] Peter Hänggi, Peter Talkner*, Michal Borkovec, Rev. Mod. Phys., 1990, 2(62): 253-254
[10] Long Gao, et al., Chinese Physics C, 2013,37(7)